\documentclass[showpacs,showkeys,amssymb,aps,prl,epsfig]{revtex4}
\newcommand{\be}{\begin{equation}} \newcommand{\ee}{\end{equation}}
\newcommand{\bea}{\begin{eqnarray}}\newcommand{\eea}{\end{eqnarray}}

\usepackage{graphicx}
\begin{document}
\preprint{SINP-TNP/02-23}
\title{Inequivalent Quantizations of the Rational Calogero Model}
\author{B. Basu-Mallick,\footnote{Email: biru@theory.saha.ernet.in} 
Pijush K. Ghosh \footnote{Email: pijush@theory.saha.ernet.in} and 
Kumar S. Gupta\footnote{Email: gupta@theory.saha.ernet.in ~ (corresponding
author)}}
\affiliation{ Theory Division\\
Saha Institute of Nuclear Physics\\
1/AF Bidhannagar, Calcutta - 700064, India.\\}

\begin{abstract}
We show that the self-adjoint extensions of the 
rational Calogero model with suitable boundary conditions leads to
inequivalent quantizations of the system. The corresponding spectrum is
non-equispaced, consisting of infinitely many positive energy states and at
most a single negative energy state. These new states appear for arbitrary
number of particles and for specific range of coupling constant.
\end{abstract}

\maketitle

The rational Calogero model is described by $N$ identical particles interacting
with each other through a long-range inverse-square and harmonic
interaction on the line \cite{calo3}. This is one of the most
celebrated examples of exactly solvable many-particle quantum mechanical
systems \cite{pr}. This model and its variants \cite{pr} are 
relevant to the study of many branches of contemporary physics, including
generalized exclusion statistics \cite{poly},
quantum hall effect \cite{qhe}, Tomonaga-Luttinger liquid \cite{ll}, quantum
chaos \cite{rmt},  quantum electric transport in mesoscopic system \cite{qet},
spin-chain models \cite{hs}, Seiberg-Witten theory \cite{sw} and
black holes \cite{black}.

The spectrum of the $N$-particle rational Calogero model was first obtained
almost three decades ago, which has since
been analyzed using a variety of different techniques \cite{brink}. In
his original work \cite{calo3}, Calogero used the boundary condition
that the wavefunction and the current vanish when any two or more particles 
coincide. With this boundary condition the Hamiltonian is self-adjoint,
which ensures the reality of eigenvalues as well as the completeness of the 
states. The central issue that we
address in this Letter is whether the spectrum obtained for rational Calogero 
model is unique or does the system admit inequivalent quantizations leading 
to different spectra? One way to address this issue is to look for
more general boundary conditions for which the Calogero Hamiltonian is
self-adjoint. The possible boundary conditions for an operator are encoded in 
the choice of its domains, which are classified by the self-adjoint extensions 
\cite{reed} of the operator. We are thus naturally led to the study of the 
self-adjoint extensions of the Calogero model. It may be noted that
self-adjoint extensions are known to play important roles in a variety of 
physical contexts
including Aharonov-Bohm effect \cite{gerbert}, two and three dimensional
delta function potentials \cite{jackiw}, anyons \cite{manuel}, anomalies
\cite{esteve}, $\zeta$-function renormalization \cite{falo}, 
particle statistics in one dimension \cite{bal} and black
holes \cite{trg}. Indeed, the self-adjoint extensions of the rational
Calogero model in absence of the confining interaction has recently
been studied \cite{we}.

In this Letter we shall show that the Calogero model in presence of the
confining interaction can indeed be consistently quantized with choices of
boundary conditions different than what was considered in Ref. \cite{calo3}. 
It will be shown that under certain conditions, the corresponding Hamiltonian 
admits self-adjoint extensions labelled by $e^{iz}$ where 
$z \in R$ (mod $2 \pi$). The parameter $z$ classifies the possible boundary
conditions for which the Hamiltonian is self-adjoint. 
In any given situation, the physical interpretation of $z$ 
depends on the details of the particular problem \cite{reed}. For example, 
in a description of black holes in terms of the Calogero model \cite{black}, 
$z$ is related to the mass and entropy of the black hole \cite{trg}. 
As another example, in the realization of generalized exclusion statistics
within the framework of Calogero model \cite{poly}, $z$ is related to the 
statistical parameter. 

As a consequence of the self-adjoint extensions, we get a new 
class of bound states for the Calogero model. Unlike the known spectrum
obtained by Calogero \cite{calo3}, we find infinitely many
energy states which are not equispaced except for special
values of $z$. Moreover, the spectrum in general includes a single 
negative energy bound state. This is the first time in the 
literature that the existence of non-equispaced energy levels 
with a single negative energy state have been found for the 
the rational Calogero model. The spectrum depends explicitly on the value
of $z$, leading to inequivalent quantizations of this system.

The Hamiltonian of the rational Calogero model is given by
\be
H = - \sum^{N}_{i=1} \frac{{\partial}^2}{\partial x_i^2} +
\sum_{i \neq j} \left [ \frac{a^2 - \frac{1}{4}}{(x_i - x_j)^2} +
\frac{\Omega^2}{16} (x_i - x_j)^2 \right ]
\label{e0}
\ee
where $a$, $\Omega$ are constants, 
$x_i$ is the coordinate of the $i^{\rm th}$ particle and
units have been chosen such that $2 m {\hbar}^{- 2} = 1$.
We are interested in finding normalizable solutions of the
eigenvalue problem
\be
H \psi = E \psi.
\label{e1}
\ee
Following \cite{calo3}, we consider the above eigenvalue equation
in a sector of configuration
space corresponding to a definite ordering of particles given by
$x_1 \geq x_2 \geq
\cdots \geq x_N$. The translation-invariant
 eigenfunctions of the Hamiltonian $H$ can be written as
\be
\psi = \prod_{i <j} \left (x_i - x_j \right )^{a + \frac{1}{2}} \
\phi (r) \ P_k (x),
\label{e2}
\ee  
where $x \equiv (x_1, x_2, \dots, x_N)$,
\be 
r^2 = \frac{1}{N} \sum_{i < j} (x_i - x_j)^2 \ \
\label{e3}
\ee
and $P_k (x)$ is a translation-invariant as well as  homogeneous 
polynomial of degree $k(\geq 0)$ which satisfies the
equation
\be
\left[ \sum^{N}_{i=1}\frac{{\partial}^2}{\partial x_i^2}
+ \sum_{i \neq j} \frac{ 2 (a + \frac{1}{2}) }{(x_i -
x_j)}  \frac{{\partial}}{\partial x_i} 
\right] P_k (x) = 0.
\label{e4}
\ee
The existence of 
complete solutions of (\ref{e4}) has been discussed by Calogero
\cite{calo3}.
Substituting Eqn. (\ref{e2}) in Eqn. (\ref{e1}) and using Eqns. (\ref{e3}-
\ref{e4}) we get
\be
\tilde{H} \phi
= E \phi,
\label{e5}
\ee  
where
\be
\tilde{H} = \left [ - \frac{d^2}{dr^2} - (1 + 2 \nu )
\frac{1}{r} \frac{d}{d r} + w^2 r^2 \right ] 
\label{e9}
\ee
with $w^2 = \frac{1}{8} \Omega^2 N $ and
\be  
\nu = k + \frac{1}{2}(N - 3) + \frac{1}{2} N (N-1)(a + \frac{1}{2}).
\label{e6}
\ee
$\tilde{H}$ is the effective Hamiltonian in the ``radial'' direction.
Following \cite{we}, it can be easily shown that $\phi(r) \in L^2[R^+,d\mu]$
where the measure is given by $d\mu = r^{1 + 2 \nu} dr$.

The Hamiltonian $\tilde{H}$ is a symmetric (Hermitian) operator on the domain
$D(\tilde{H}) \equiv \{\phi (0) = \phi^{\prime} (0) = 0,~
\phi,~ \phi^{\prime}~  {\rm absolutely~ continuous} \} $. 
To determine whether 
$\tilde{H}$ is self-adjoint \cite{reed} in $D(\tilde{H})$, 
we have to first look for square integrable solutions of the equations 
\be
\tilde{H^*} \phi_{\pm} = \pm i \phi_{\pm},
\label{e10}
\ee
where $\tilde{H^*}$ is the adjoint of $\tilde{H}$ (note that $\tilde{H^*}$
is given by the same differential operator as $\tilde{H}$ although their
domains might be different). 
Let $n_+(n_-)$ be the total number of square-integrable, independent solutions 
of (\ref{e10})
with the upper (lower) sign in the right hand side. Now $\tilde{H}$ 
falls in one of the following categories \cite{reed} :\\
1) $\tilde{H}$ is (essentially) self-adjoint iff
$( n_+ , n_- ) = (0,0)$.\\
2) $\tilde{H}$ has self-adjoint extensions iff $n_+ = n_- \neq 0$.\\
3) If $n_+ \neq n_-$, then $\tilde{H}$ has no self-adjoint extensions.\\

The solutions of Eqn. (\ref{e10}) are given by
\be
\phi_{\pm} (r) = {\mathrm e}^{- \frac{w r^2}{2}} 
U \left ( d_\pm, c, w r^2 \right ),
\label{e11}
\ee
where $d_{\pm} = \frac{1+ \nu }{2} \mp \frac{i}{4 w}$, $c = 1+ \nu $ and  $U$
denotes the confluent hypergeometric function of the
second kind \cite{abr}. The asymptotic behaviour of $U$ \cite{abr} together
with the exponential factor in Eqn. (\ref{e11}) ensures that 
$\phi_{\pm} (r)$ vanish at infinity. The solution in Eqn. (10) have
different short distance behaviour for $\nu \neq 0$ and $\nu = 0$. 
From now onwards, we shall restrict our discussion to the case for 
$\nu \neq 0$, the analysis for $\nu = 0$ being similar. When $\nu \neq 0$, 
$U(d_{\pm},c,w r^2)$ can be written as 
\be
U \left ( d_{\pm},c,w r^2 \right )=
C
\bigg [ \frac{M \left ( d_{\pm}, c, w r^2 \right )}
{\Gamma (b_{\pm}) \Gamma (c)}
 -  \left ( w r^2 \right )^{1 -c}
\frac{M \left ( b_{\pm}, 2-c, w r^2 \right )}
{\Gamma (d_{\pm}) \Gamma (2-c)} \bigg ],
\label{e12}
\ee
where $b_\pm = \frac{1-\nu}{2} \mp \frac{i}{4w}$, $C = \frac{\pi}{{\mathrm
sin} (\pi + \nu \pi)}$ and 
$M$ denotes the confluent hypergeometric function of the first kind
\cite{abr}.
In the limit $r \rightarrow 0$, $M(d_{\pm},c,wr^2) \rightarrow 1$. 
This together with Eqns. (\ref{e11}) and (\ref{e12}) implies that as
$r \rightarrow 0$,
\be
|\phi_{\pm} (r)|^2 d \mu \rightarrow
\left [ A_1 r^{(1 + 2 \nu )} + A_2 r + A_3 r^{(1 - 2 \nu )} \right ] dr,
\label{e14}
\ee
where $A_1, A_2$ and $A_3$ are constants independent of $r$. 
From Eqn. (\ref{e14}) it is
now clear that in the limit $r \rightarrow 0$, the functions
$\phi_{\pm} (r)$ are not square-integrable if ${\mid \nu \mid} \geq 1 $.
In that case, $n_+ = n_- = 0$ and $\tilde{H}$ is essentially self-adjoint
in the domain $D(\tilde{H})$. However, if either $ 0 < \nu < 1$ or
$ -1 < \nu < 0$, the functions $\phi_{\pm} (r)$ are indeed square-integrable.
Thus if $\nu$ lies in these ranges, we have $n_+ = n_- = 1$ and 
Hamiltonian $\tilde{H}$ is not self-adjoint in $D(\tilde{H})$ but
admits self-adjoint extensions.
The domain $D_z(\tilde{H})$ in which $\tilde{H}$ is self-adjoint contains all
the elements of $D(\tilde{H})$ together with elements of the form
$\phi_+ + {\mathrm e}^{iz} \phi_-$, where $ z \in R$ (mod $2 \pi$) \cite{reed}.
We can similarly show that $n_+ = n_- = 1$ for $\nu = 0$ as well. Thus the
self-adjoint extensions of this model exist when $-1 < \nu < 1$. It may be
noted that the values of $n_+$ and $n_-$ as well as the allowed range of
$\nu$ obtained above is the same as that found in Ref. \cite{we}, which
discussed the Calogero model without the confining term. In both these
cases,
the existence of the self-adjoint extension is essentially determined by the 
nature of the singularity at $r=0$. However, the domain 
$D_z(\tilde{H})$ obtained above is very different from the corresponding
domain found in Ref. \cite{we}. This is due to the fact that the 
presence of the confining
potential affects the expressions of $\phi_{\pm}(r)$, which in turn determine
the allowed domain of $\tilde{H}$. As discussed below,
this difference in the structure of the domains leads to a completely
different spectrum in presence of the confining potential.

The range of $\nu$ required for the existence of the self-adjoint extension 
together with Eqn. (8) implies that for given
values of $N$ and $k$, $a + \frac{1}{2}$ must lie on the range
\be
- \frac{ N - 1 + 2 k}{N ( N-1)} < a + \frac{1}{2} < -
\frac{ N - 5 + 2 k}{N ( N-1)}.
\label{pp}
\ee
For $N \geq 3$, we have the following 
classifications of the boundary conditions depending on the value of the
parameter $a+ \frac{1}{2}$.\\
(i) $a + \frac{1}{2}\geq \frac{1}{2}$ : This corresponds to the boundary 
condition considered  by Calogero for 
which both the wave-function and the current vanish as $x_i \rightarrow x_j$.
In this case, $\nu > 1$ for all values of $k \geq 0$.
The corresponding Hamiltonian is essentially self-adjoint in the domain
$D(\tilde{H})$, leading to a unique quantum theory.\\
(ii) $ 0 < a + \frac{1}{2} <  \frac{1}{2}$ :  
The wave-function vanishes in the limit $x_i \rightarrow x_j$, though
the current may show a divergent behaviour in the same limit. Such a boundary
condition on the wave-function is quite similar to what one encounters for
strongly repulsive $\delta$-function Bose gas. In this case 
$\nu$ is positive and $k$ must be equal to zero so that $\nu$ may belong to 
the range 
$0 < \nu < 1$. The corresponding constraint on $a + \frac{1}{2}$ is given by
$0 < a + \frac{1}{2} < \frac{5 - N }{N(N-1)}$, which can only be
satisfied for $N = 3$ and $4$.\\
(iii) $ -\frac{1}{2} < a + \frac{1}{2} <  0$ :
The lower bound on $a+ \frac{1}{2}$ is obtained from the condition that the
wavefunction be square-integrable. The parameter $a + \frac{1}{2} $ in this
range leads to a singularity in the wavefunction resulting from the
coincidence of any two or more particles. 
Using permutation symmetry, such an eigenfunction can be extended to the
whole of configuration space, although not in a smooth fashion.
The new quantum states in this case exist for arbitrary $N$ and even for
non-zero values of $k$. In fact,
imposing the condition that the upper bound on $a + \frac{1}{2}$  should be
greater than $-\frac{1}{2}$, we find from Eqn. (13) that $k$ is restricted as 
$k < \frac{1}{4} \left ( N^2 - 3 N + 10 \right )$. It can also be
shown that there are only two allowed values of $k$ when both $N$ and 
$a + \frac{1}{2}$ are kept fixed.

In order to determine the spectrum 
we note that the solution to Eqn. (\ref{e5}) which is
bounded at infinity is given by 
\be
\phi (r) = B {\mathrm e}^{- \frac{w r^2}{2}} 
U(d,c,w r^2),
\label{e15}
\ee
where $d = \frac{ 1+ \nu }{2} - \frac{E}{4 w}$ and $B$ is a constant.
In the limit $r \rightarrow 0$, 
\be
\phi (r) \rightarrow 
B C \left [ \frac{1}{\Gamma (b) \Gamma (c)}  
- \frac{w ^{-\nu}  r^{-2 \nu }}{\Gamma (d) \Gamma (2-c)} \right ],
\label{e16}
\ee
where $b = \frac{1 - \nu}{2} - \frac{E}{4w}$.
On the other hand, as $r \rightarrow 0$,
\be
\phi_+ + {\mathrm e}^{iz} \phi_-  \rightarrow  C \bigg [
\frac{1}{\Gamma (c)}  \left (
\frac{1}{\Gamma (b_+) }
+\frac{{\mathrm e}^{iz}}{\Gamma (b_- ) } \right )
  -  \frac{ w ^{-\nu} r^{-2 \nu }}{\Gamma (2-c)}   
\left ( \frac{1}{\Gamma (d_+) } 
+\frac{{\mathrm e}^{iz}}{\Gamma (d_-) } \right ) \bigg ].
\label{e17}
\ee
If $\phi (r) \in D_z(\tilde{H})$, then the coefficients of different powers of
$r$ in Eqns. (\ref{e16}) and (\ref{e17}) must match. Comparing the coefficients 
of the constant term and $r^{-2 \nu }$ in 
Eqns. (\ref{e16}) and (\ref{e17}) we get
\be
f(E) \equiv \frac{\Gamma \left ( \frac{ 1 - \nu }{2} - \frac{E}{4 w} \right )}
{\Gamma \left (\frac{ 1 + \nu }{2} - \frac{E}{4 w} \right ) } =
\frac{\xi_2 {\mathrm cos}(\frac{z}{2} - \eta_1)}
{\xi_1 {\mathrm cos}(\frac{z}{2} - \eta_2)},
\label{e18}
\ee 
where $\Gamma \left ( \frac{ 1 + \nu }{2} + \frac{i}{4 w} \right )
\equiv \xi_1 {\mathrm e}^{i \eta_1}$
and
$\Gamma \left ( \frac{ 1 - \nu }{2} + \frac{i}{4 w} \right )
\equiv \xi_2 {\mathrm e}^{i \eta_2}$.
For given values of the parameters $\nu$ and  $w$, 
the bound state energy $E$ is obtained from Eqn. (\ref{e18}) as a function of
$z$. The corresponding eigenfunctions are
obtained by substituting $\phi(r)$ from Eqn. (\ref{e15}) into Eqn. (3).
Different choices of $z$ thus leads to 
inequivalent quantizations of the many-body Calogero model.
Moreover from Eqn. (\ref{e18}) we see that for fixed value of $z$, 
the Calogero model with 
parameters $(w, \nu)$ and $(w,-\nu)$ produces identical energy 
spectrum although the corresponding wavefunctions are different.

\begin{figure}
\begin{center}
\includegraphics[width=7cm]{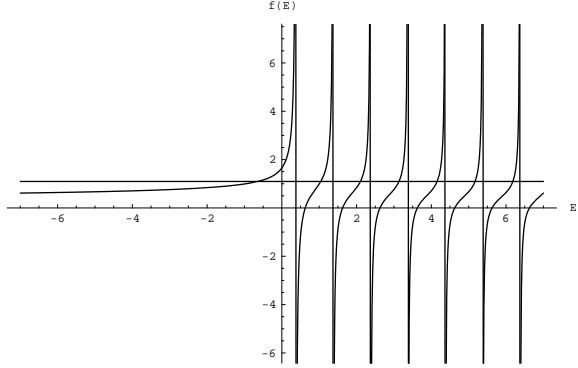}
\end{center}
\caption { \label{fig1} A plot of Eqn. (\ref{e18}) using Mathematica
with $w = 0.25$, $\nu = 0.25 $ and 
$z = -1.5$. The horizontal straight line corresponds the value of the r.h.s of
Eqn. (\ref{e18}).} 
\end{figure}

The following features about the spectrum may be noted: \\
1) We have obtained the spectrum analytically when the r.h.s.
of Eqn. (\ref{e18})
is either 0 or $\infty$. When the
r.h.s. of Eqn. (\ref{e18}) is $0$, we must have the situation where 
$\Gamma \left (\frac{ 1 + \nu }{2} - \frac{E}{4 w} \right )$ blows up, i.e.
$E_n = 2 w ( 2 n + \nu + 1)$ where $n$ is a positive integer. 
This happens for the special choice of $z = z_1 = \pi + 2 \eta_1$.
These
eigenvalues and the corresponding eigenfunctions are analogous to those
found by Calogero although for a different parameter range. 
Similarly, when the r.h.s. of Eqn. (\ref{e18}) is $\infty$, 
an analysis similar
to the one above shows that $E_n  = 2 w ( 2 n - \nu + 1)$. 
This happens for the special value of $z$ given by 
$z = z_2= \pi + 2 \eta_2$. \\
2) For choices of $z$ other than $z_1$ or $z_2$, the nature of the spectrum
can be understood from Figure 1, which is a plot of Eqn. (\ref{e18}) for
specific values of $\nu, z$ and $w$.
In that plot, the horizontal straight line  corresponds to the 
r.h.s of Eqn. (\ref{e18}). The energy eigenvalues are obtained from the
intersection of $f(E)$ with the horizontal straight line.
Note that the spectrum
generically consists of infinite number of positive energy solutions and at
most one negative energy solution. The existence of the negative energy
states can be understood in the following way. For large negative values of
$E$, the asymptotic value of $f(E)$ is given by $(\frac{E}{4w})^{- \nu}$
\cite{abr}, which monotonically tends to 0 or $+ \infty$ 
for $\nu > 0$ or $\nu < 0$ respectively. When $\nu > 0$, the negative energy
state will exist provided r.h.s. of Eqn. (\ref{e18}) lies between 0 and 
$\frac{\Gamma  ( \frac{ 1 - \nu }{2})}
{\Gamma  (\frac{ 1 + \nu }{2} )}$. Similarly, when $\nu < 0$, the 
negative energy state will exist when the r.h.s. of Eqn. (\ref{e18})
lies between 
$\frac{\Gamma  ( \frac{ 1 - \nu }{2})}
{\Gamma (\frac{ 1 + \nu }{2} ) }$ and $+ \infty$.
For any given values of $\nu$ and $w$, the 
position of the horizontal straight line in Fig. 1 can always be adjusted 
to lie anywhere between $-\infty$ and $+\infty$ by 
suitable choices of $z$. Thus the spectrum would always contain a negative
energy state for some choice of the parameter $z$.    \\
3) Contrary to the spectrum of the rational Calogero model, 
the energy spectrum obtained from Eqn. (\ref{e18}) is not equispaced for
finite
values of $E$ and for generic values of $z$. For example, it is seen from
Eqn. (\ref{e18}) that the ratio
\be
\frac{f(E + 4w)}{f(E)} = \frac{\frac{E}{4w} + \frac{1 - \nu}{2}}
                         {\frac{E}{4w} + \frac{1 + \nu}{2}}
\ee
in general is not unity except when $E \rightarrow \infty$. 
This may seem surprising with the presence of $SU(1,1)$ as
the spectrum generating algebra in this system \cite{fub}, which demands 
that the eigenvalues be evenly spaced. In order to address this issue, we   
consider the action of the dilatation generator $D = \frac{1}{2} \left ( r  
\frac{d}{dr} + \frac{d}{dr}r \right )$ on an element 
$\phi(r) = \phi_+(r) + {\mathrm e}^{iz} \phi_-(r)$.  
In the limit $r \rightarrow 0$, we have
\be
D \phi = \frac{C}{2} \bigg [
\frac{1}
{\Gamma (c)} \left (
\frac{1}{\Gamma (b_+) }
+\frac{{\mathrm e}^{iz}}{\Gamma (b_-) } \right )
  - \frac{r^{-2 \nu } (1 - 4 \nu) }
{\Gamma (2-c)}
\left ( \frac{1}{\Gamma (d_+) }
+\frac{{\mathrm e}^{iz}}{\Gamma (d_-) } \right ) \bigg ].
\label{e19}
\ee
We therefore see that $D \phi (r) \in D_z (\tilde{H})$ only for $z = z_1$ or
$z= z_2$.
Thus the generator of dilatations does not in
general leave the domain of the Hamiltonian invariant
\cite{dh,esteve,jackiw,we}. Consequently, $SU(1,1)$ cannot be implemented as
the spectrum generating algebra except for $z = z_1$, $z_2$.\\
4) For $N \geq 3$, the range of $a + \frac{1}{2}$  for which the 
new quantum states have been found is different from what was used in 
Ref. \cite{calo3}. 
The $N=2$ Calogero model however admits new quantum states even in 
the range of $a + \frac{1}{2}$ considered in Ref. \cite{calo3}.
When $N=2$, $k$ must be equal to
zero and Eqn. (8) gives $\nu = a$. In this case, 
the system therefore admits self-adjoint
extensions and new quantum states when $-1 < a < 1$. It may be noted that the
eigenvalue problem for $N=2$ was solved in Ref.\cite{calo3} with the 
condition that $a > 0$. Thus when $0 < a < 1$,
our analysis predicts a larger family of solutions labelled by the
parameter $z$. This set of solutions reduces to that
found in Ref. \cite{calo3} for $z = z_1$.\\
5) It may be interesting to compare the spectrum obtained above with that
found in Ref. \cite{we}, where the 
self-adjoint extension of the Calogero model
without the confining term was discussed. In the latter case, the spectrum
consists of at most
one negative energy bound state and infinite number of scattering states
with momentum dependent phase shifts. In the presence of the confining
potential, as discussed above, we get at most one negative energy bound
state and an infinite number of positive energy bound states which are in
general not equispaced. It may also be noted that the spectrum found in Ref.
\cite{we} cannot be obtained as the $w \rightarrow 0$ limit of that
obtained in this paper. This is due to the fact that Eqn. (17), which
determines the spectrum in the present case, becomes singular in the 
$w \rightarrow 0$ limit.   

In conclusion, we have presented a new quantization scheme for the rational
Calogero model. The non-equispaced nature of the energy levels and the 
existence of a negative energy bound state are some of the salient features
that emerge from our analysis. It is expected that the generaized exclusion
statistics parameter\cite{poly} of this model would be a function of both
$z$ and $\nu$, since the energy spectrum depends on these parameters.
We can ascertain this for the special case of $z=z_1$ and $z_2$,
when Eq. (\ref{e18}) can be solved exactly. 
The generalized exclusion statistics is believed to
play an important role in one dimensional non-fermi liquids as well as in
the edge excitations in the fractional quantum Hall effect.
Thus, it would be interesting to
investigate the generalized exclusion statistics and the thermodynamic
properties of Calogero models for arbitrary $z$ with the spectrum as
described here.

\noindent
\acknowledgments{ The work of PKG is supported(DO No. SR/FTP/PS-06/2001) by
the SERC, DST, Govt. of India, under the Fast Track Scheme for Young
Scientists:2001-2002. }

\end{document}